\documentclass[%
 reprint,
 amsmath,amssymb,
 aps,
pra,
]{revtex4-2}

\usepackage{graphicx}
\usepackage{dcolumn}
\usepackage{bm}
\usepackage[colorlinks=true, urlcolor=blue, linkcolor=blue, citecolor=blue]{hyperref}

\begin{document}

\preprint{APS/123-QED}

\title{Lifting a granular box by a half-buried rod}

\author{Ting-Heng Hsieh and Tzay-Ming Hong }
\thanks{ming@phys.nthu.edu.tw}
\affiliation{Department of Physics, National Tsing Hua University, Hsinchu, Taiwan 30013, Republic of China}\
\date{\today}

\begin{abstract}
We studied an interesting experiment that showed a half-buried chopstick lifting a full bottle of granules off the table. In Janssen's theory, the friction force provided by the container wall helps alleviate the weight of the granules. How can a thin rod with a much less contact area support the full weight plus that of the container? Insights are gained by allowing the friction on the wall to change direction before solving the Janssen equation. We obtained the analytic expression for the critical depth of granules that enables a successful lift-off. In addition, we established that the stick-slip phenomenon exists during a failed lift-off by analyzing the frequency of fluctuations in the pull force. Finally, a photoelasticity experiment was employed to directly visualize the stress field sensitive to the pull force, and verify the directional change of friction force from the wall.
\end{abstract}

\maketitle

\section{Introduction}

The pressure in a liquid or solid of mass density $\rho$ is proportional to the depth $h$ as $\rho g h$ where $g$ is the gravitational constant. Imagine pouring salt from the shaker, the granules at the uppermost layers flow like a liquid, while the rest of the salt remains solid-like. So it is interesting to ask how the pressure distributes in granules that exhibit either or both phases. The answer depends crucially on whether the granules are stored in a container. If not, $\rho g h$ remains valid, as for a sandpile\cite{On_the_stress_depression_under_a_sandpile,Searching_for_the_Sand-Pile_Pressure_Dip,Stress_distribution_in_a_sandpile_formed_on_a_deected_base,particle_aspect_ratio_on_pressure_of_sandpiles,Finite_element_simulation_of_the_pressure_dip_in_sandpiles,Continuum_Theories_Previously_Failed_Describe_Sandpile_Formation, The_sandpile_revisited,An_investigation_pressure_dip_phenomenon}. Otherwise, a phenomenon called the Janssen effect \cite{Jassen_first,Granular_solids_liquids_and_gases, The_effect_of_random_friction, Stress_in_silos,Dynamical_Janssen_Effect_on_Granular_Packing_with_Moving_Walls,Discrete_element_simulations_of_stress_distributions_in_silos,Translation_janssen,Local_stresses_in_the_Janssen_granular_column,Stress_distribution_in_two-dimensional_silos,Janssen_effect_in_dynamic_particulate_systems,Magnetic_Janssen_effect} will enter. The pressure inside a salt shaker or grain silo initially increases as  $\rho g h$, but saturates exponentially when $h$ exceeds a certain length scale, depending on the size and frictional coefficient of the container. The deviation from $\rho g h$ comes from the frictional force exerted by the inside wall of the container, which lifts and eventually cancels the weight of the granules at the same layer exactly. It has been reported in some 
special cases, such as purposely pulling down the wall of the container relative to the granules or using a container with a small radius \cite{Reverse_janssen}, exponential growth with $h$ can be elicited. This is dubbed the reversed Janssen effect.

Janssen's theory is based on three key assumptions: (1) the pressure is isotropic and not sensitive to its horizontal position, (2) the friction force from the inside wall of the container equals $\mu N$ where $\mu$ is proportional to the coefficient of static friction by some fraction $k<1$ and $N$ is the normal force, and (3) the density of the granules is evenly distributed in the container. It is easy to point out the shortcomings of these assumptions. But what amazes us is that the predictions by Janssen's theory are successful in most cases. So our strategy is to follow Janssen's theory while bearing in mind to be critical and take the opportunity to update Janssen's theory when major discrepancies occur.

Even when filled to the same height and packing ratio, the detailed arrangement of granules can never be repeated. This implies a huge degeneracy for the configuration of jammed granules. Consequently, the distribution of internal pressure or force chains depends sensitively on the history, for instance, (1) whether the system was prepared by pouring the granules from the center of the container or by sprinkling them uniformly and (2) whether the container wall was pulled purposely upwards or downwards. To get rid of these artifacts, we see to it that the granular samples are shaken or/and compressed thoroughly before each run of the experiment. 

With wide applications in various projects such as large rockfill dams, roadbed filling, ballast, and airport embankments, granular soils refer to special materials that exhibit high strength, easy availability, permeability, and compaction characteristics\cite{Cohesionless_Materials_for_Rockfill_Dams,Settlement_time_behaviour,railway_ballast_vibrations,Shape_Properties_of_Railroad_Ballast,Embankment_on_Soft_Deposit,Grain_size_and_time_effect,loess_roadbed_filling}. In contrast, the common soil is not strictly granules. But for the sake of argument, imagine the imminent issue of water and soil conservation in the era of global climate change. It is of uttermost importance to study how plants help prevent the mud-and-stone flow, and what kind of root system is most effective. This motivated us to answer three questions: (1) how does the existence of a half-buried rod, simulating the root of plants, affect the pressure distribution of granules or Janssen's theory? (2) how does the answer to the previous question vary if the rod is being pulled by different magnitudes of force? (3) is it possible to lift the granules along with the container above the ground? 

Intuitively, the answer to question (3) is a clear-sounding no since the surface area of the rod is much smaller than that of the container, while the latter can only alleviate the granular weight partially.To our knowledge, Furuta {\it et al.} \cite{Packing-dependent_granular_friction_exerted_on_rod} was the first to study the granular system with a buried rod as recently as  2019. Interested in how the packing density and granular size affect the friction force on a withdrawn rod, they reported evidence of shear-induced solidification at a high packing fraction. We were drawn to this interesting setup for a less academic reason, i.e., we overheard that some YouTuber \cite{youtube1} managed to lift a Yalult bottle full of rice with a buried bamboo chopstick. This motivated us to not only verify the plausibility of this feat and find the critical depth, but also illuminate how the pressure distribution is modified by the presence of the rod and its pull force in such a way as to enable enough friction force on the rod to pull off the trick.

\section{Experimental setup}
\begin{figure}
	\includegraphics[width=8.6cm]{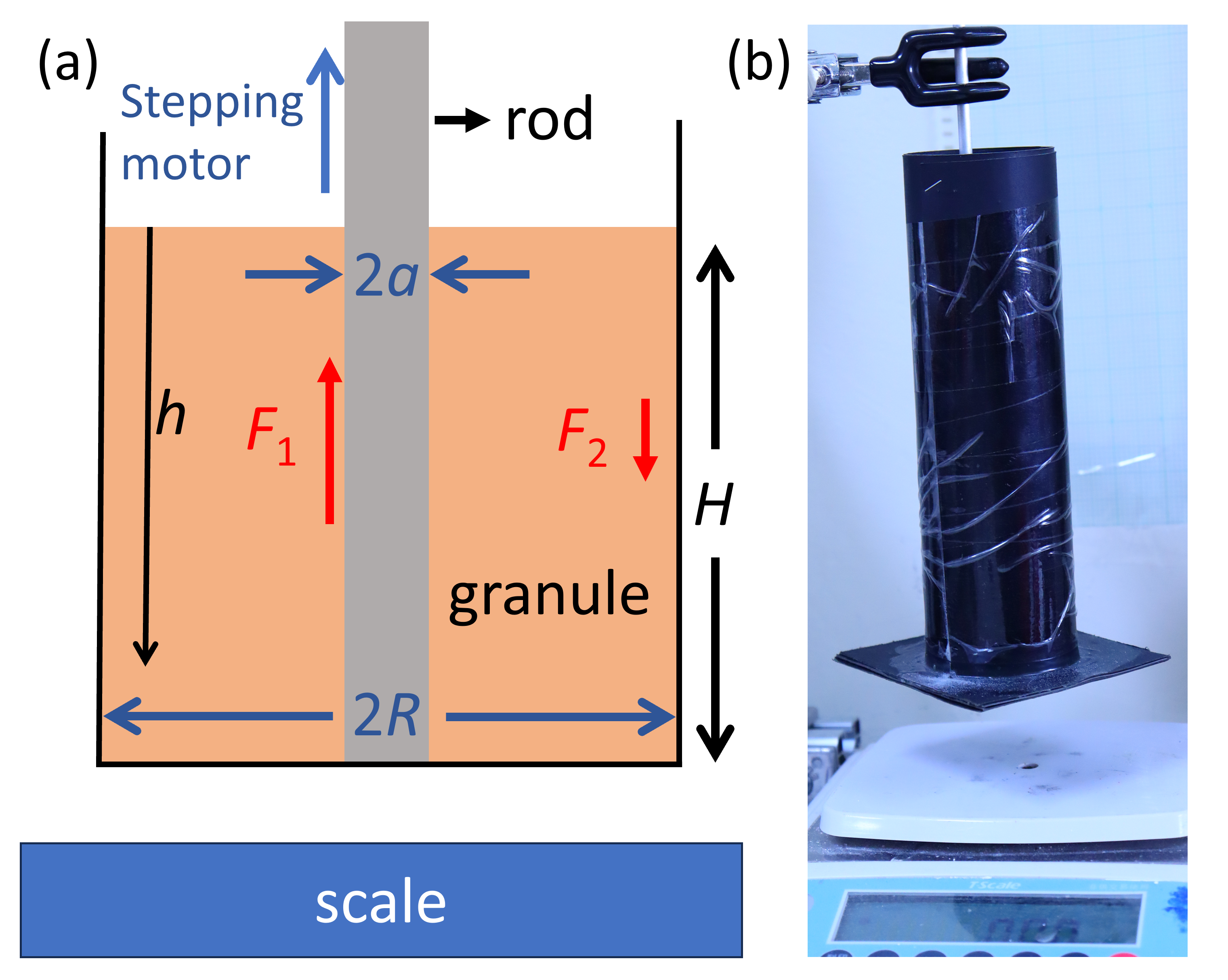}
	
    \caption{(a) Schematic experimental setup for a lifted granule system. The scale reading decreases as the rod is gradually pulled by the stepping motor. The granules will experience forces  $F_1$ and $F_2$ from rubbing with the rod and inner wall of the container. The direction of $F_2$ is expected to evolve from being initially upward to possibly downward, depending on whether the system is successfully lifted above the scale. A photo of the real setup is shown in (b). }
    \label{setupgranule}        
\end{figure}

We utilize plastic granular spheres with a roughly uniform diameter of about 0.5 mm and an aluminum rod with five different radii $a$ = 0.25$\sim$1.25 cm. Four cylindrical containers made of thick paper with radii $R$ = 1.5$\sim$4.5 cm are tailored to weight at the same mass at  $m$ = 75 g to minimize the number of control parameters. A rod is inserted vertically at the center of the container and touches its bottom before we pour in the granules. To eliminate the effect of the initial condition, i.e., how the granules are prepared, the container was shaken vertically and horizontally for tens of seconds by a vibrator of amplitude 0.45 cm and frequency 10 Hz. We ensure the granules reached the random close packing of density 0.67 before each trial. The rod is pulled up by a stepping motor with a low speed of 0.17 cm/s, as shown schematically in Fig. \ref{setupgranule}(a). An electronic scale with a resolution of 0.02 g is placed beneath the container. The critical depth $H_c$ of granules required to lift the granules and the container can then be determined when the reading in the scale goes to zero in Fig. \ref{setupgranule}(b). It will also come in handy when we try to elucidate the slip-and-stick motion experienced by the rod in the failed attempt of $H< H_c$. 

To facilitate the visualization of how the force chains evolve as we gradually crank up the pull force and how they can help us distinguish the successful as opposed to the failed feat of lifting the system, we employ the method of photoelasticity. The granules were replaced with photoelastic cylindrical plates (Vishay PS-6) with a radius of 5 mm and thickness of 3 mm and illuminated by an LED light with two overlapped polarizers with perpendicular polarization directions. Aided by the principle of birefringence, we can deduce the force chains, as will be detailed in Sec. IV. 

\section{Theory}
The friction force  $F_2$ from the inner surface of the container is normally opposite to gravity. Still, it may reverse its direction in (1) the Reverse Janssen effect by pulling down the container wall relative to the granules or using a small container radius \cite{Reverse_janssen} and (2) our rod experiment due to the countering by the upward-pointing $F_1$. Therefore, a successful reversal requires enough granules to sustain a large $F_1$. 
At the borderline of a successful lift-off, i.e., $H = H_c$, the static equilibrium requires 
\begin{equation}
    F_1= \mu_1 \int_{0}^{H_c} P(h)\cdot 2\pi a\,dh\ =\rho g \pi (R^2-a^2) H_c +mg
    \label{F1}
\end{equation}
where $\mu_1$ is the static friction coefficient between the rod and granules. Similarly,
\begin{equation}
      F_2= \mu_e \int_{0}^{H_c} P(h) \cdot 2 \pi R\,dh\ = mg+P(H_c)\pi(R^2-a^2)
      \label{F2}
\end{equation}
where the effective $\mu_e$ is smaller than the real static coefficient between granules and the inner wall. The last term signifies the normal force from the bottom of the container. Note that Eq. (\ref{F2}) is much smaller than Eq. (\ref{F1}) in our experiment. 

Mimicking the original approach by Janssen by adding the friction force from the rod, we can write down 
\begin{equation}
\begin{split}
    &\Big{(}P(h)-P(h+dh)\Big{)}\pi (R^2 -a^2) +\rho g \pi (R^2-a^2) dh \\
    &= P(h)\cdot 2 \pi(\mu_1 a- \mu_e R) dh.
    \label{suc}
\end{split}
\end{equation}
from the static equilibrium for the layer of granules between depth $h$ and $h+dh$. Note that the knowledge of a reversed, namely pointing downward, $F_2$ has been implemented by assigning an opposite sign to $F_1$.
The resulting differential equation can be readily solved to give
\begin{equation}
    P(h) = \frac{\rho g (R^2-a^2)}{2 (\mu_1 a- \mu_e R)}\Big{\{}1-\exp\big[{-\frac{2(\mu_1 a -\mu_e R)h}{R^2-a^2}} \big]\Big{\}}.
    \label{pressure}
\end{equation}
Plugging Eq. (\ref{pressure}) in Eq. (\ref{F1}) gives:
\begin{equation}
\begin{split}
    F_1 &= \mu_1 a \rho g \pi A \Big{[} H_c- \frac{A}{2}\big{(}1-e^{-\frac{2 H_c}{A}}\big{)}\Big{]}\\
    & =\rho g \pi(R^2 -a^2) H_c +mg
    \label{EQF_1}
    \end{split}
\end{equation}
where $A$ denotes $(R^2-a^2)/(\mu_1 a- \mu_e R)$. 
Combining with the experimental fact that $R \gg a$ and $H_c$ is deep enough to warrant the saturation of pressure, i.e., the exponential term can be omitted,  Eq. (\ref{EQF_1}) can be simplified to 
\begin{equation}
    \rho R^4 ( H_c- \frac{R^2}{2 \mu_1 a}) = m H_c \mu_1 a.
\end{equation}
A simple reshuffle then enables us to obtain a precise and analytic expression for the critical height:
\begin{equation}
    H_c = \frac{\rho R^6}{2 \mu_1 a (\rho R^4 -m \mu_1 a)}.
    \label{EQHC}
\end{equation}
This prediction aligns with our intuition that only a small $H_c$ is required for a large $a$ and a small $R$.
When the total granular mass is much bigger than $m$, Eq. (\ref{EQHC}) can be simplified to predict a scaling relation between the dimensionless parameters $H_c/R$ and $a/R$: 
\begin{equation}
    \frac{H_c}{R} = \frac{R}{2 \mu_1 a}.
    \label{EQ_master}
\end{equation}
Note that the phenomenological $\mu_e$ of Eq. \eqref{F2} can be determined by plugging Eq. (\ref{pressure}) in Eq. (\ref{F2}).

\section{Experimental results}
\subsection{Critical depth}

\begin{figure}
\includegraphics[width=8.6cm]{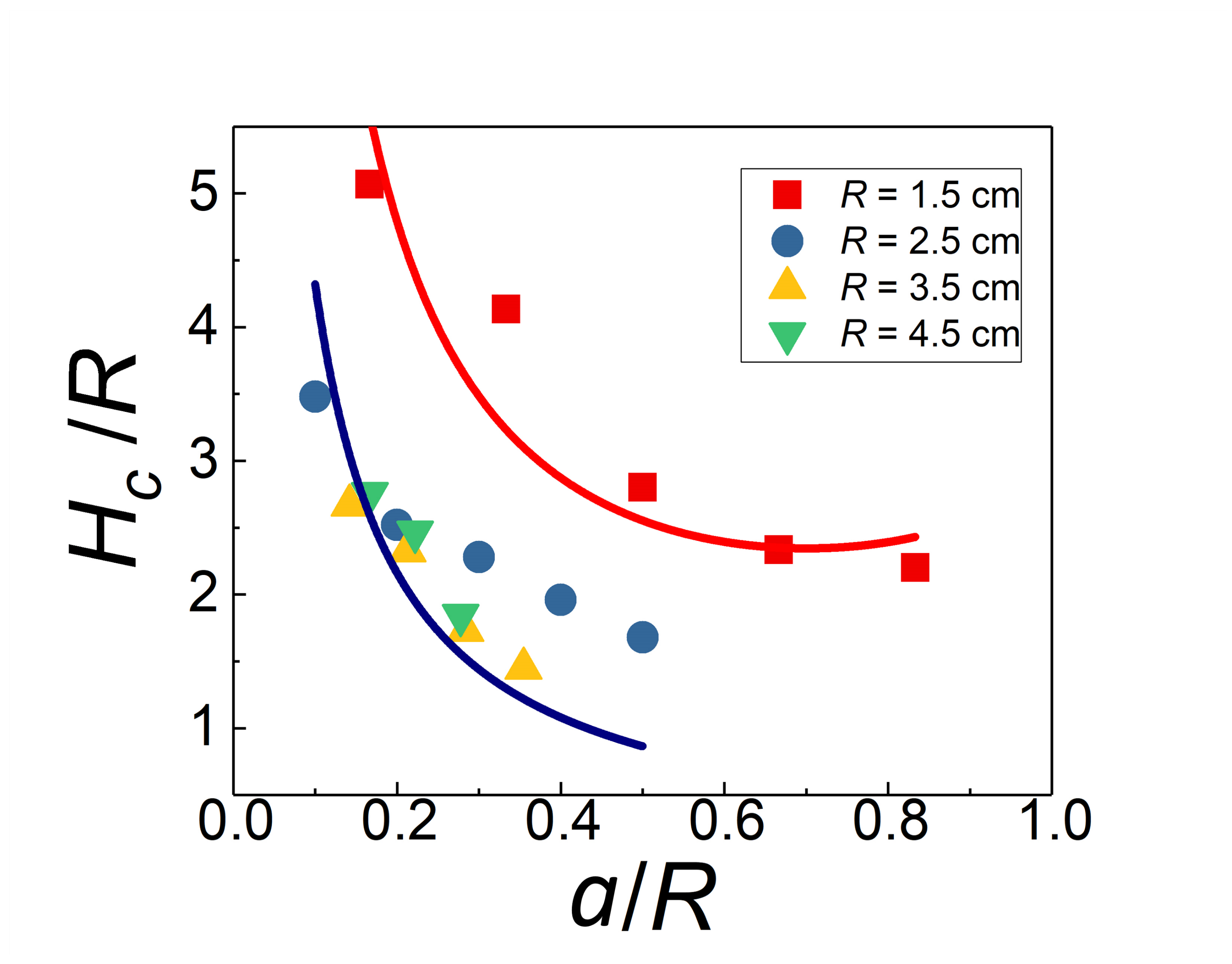}
    \caption{The rescaled critical depth $H_c$ versus the rod radius $a$. The red solid line denotes the curve fitting from Eq. (\ref{EQHC}), while the blue line is from Eq. (\ref{EQ_master}).}
    \label{criticalh}        
\end{figure}

For $R$ = 3.5 and 4.5 cm, the rescaled parameters align well with the blue master curve obeying Eq. (\ref{EQ_master}) in Fig. \ref{criticalh} with an R-square exceeding 0.80. Blue-circle data deviate considerably from such a fitting because, like the red squares, their corresponding $R$ = 2.5 cm is too small to warrant the approximation led up to Eq. (\ref{EQ_master}). They enjoy an equally high R-square when fit by Eq. (\ref{EQHC}).

\subsection{Stick-slip phenomenon}

\begin{figure}
	\includegraphics[width=6cm]{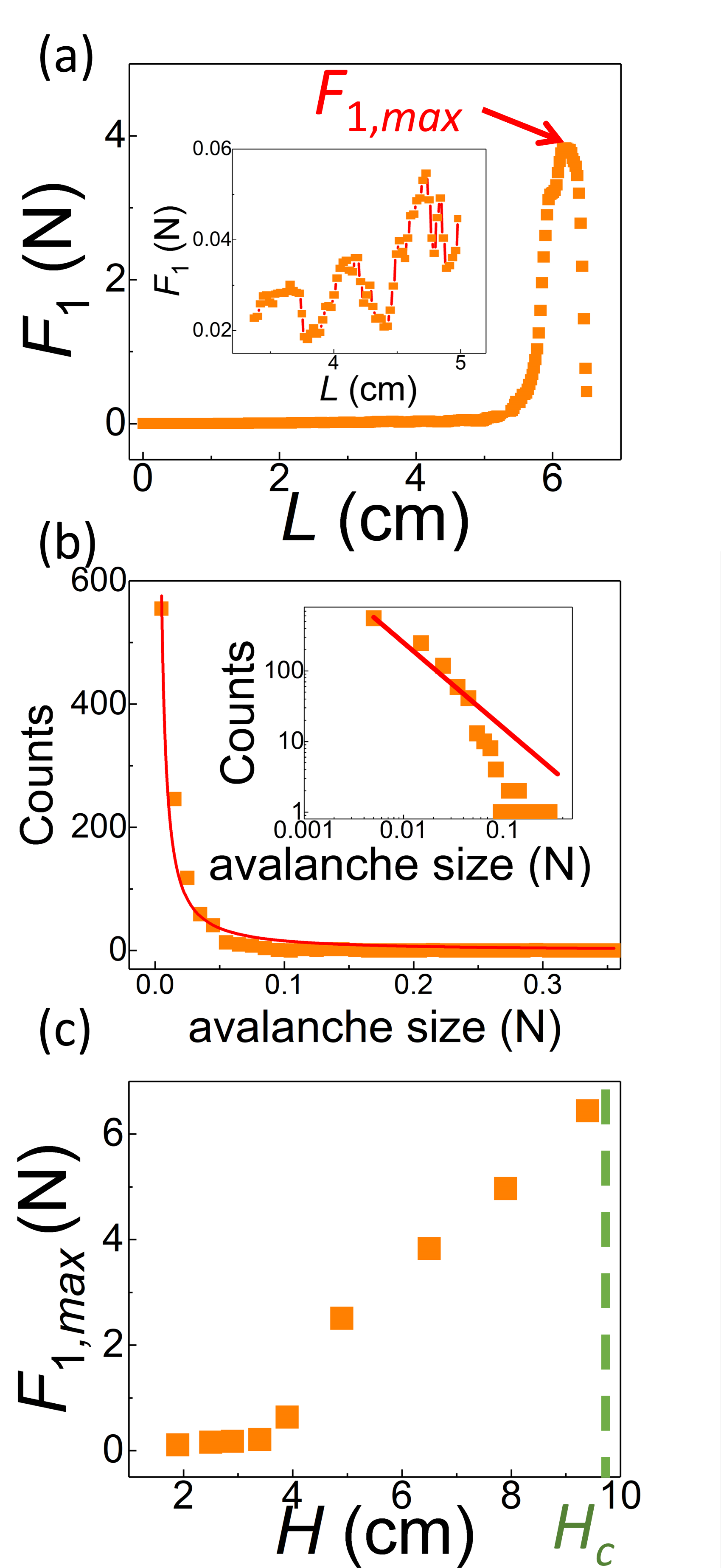}
	
    \caption{ (a) The $F_1$ versus $L$ that denotes the remaining length of the rod in granules for $R$ = 3.5 cm, $a$ = 0.5 cm, and $H$ = 6.3 cm $< H_c$ = 9.6 cm. The inset highlights the fluctuations characteristic of the stick-slip phenomenon. (b) The count and size of avalanches are fit by the orange line representing a power law with an exponent = -1.2, as vindicated by the full-log inset plot.  (c) The peak value $F_{\rm 1,max}$ is plotted for different $H$ where the fitting line follows the prediction of Eq. (\ref{EQF_1}). }
    \label{Figslip}        
\end{figure}

During an unsuccessful lift-off with $H < H_c$, the rod experiences dynamic friction with a decreasing contact area as it is gradually pulled out of the granules. This explains the reduction of $F_1$ in Fig. \ref{Figslip}(a) after the initial peak at $F_{1,{\rm max}}$ when the friction transits from static to dynamic. 
The static to dynamic friction coefficients ratio is estimated by dividing $F_{1,{\rm max}}$ by the plateau value to be roughly 4/0.4=10, an order of magnitude larger than its conventional value between two solids. We believe this gives rise to the rearrangement and loosening of granules near the rod as they collapse to fill the void left behind under the elevating rod. This explains the sharp drop after $F_{1,{\rm max}}$ without further peaks in Fig. \ref{Figslip}(a). 

Multiple fluctuations become visible in the plateau regime under scrutiny, reminiscent of the stick-slip phenomenon. To substantiate this analogy, it is necessary to check whether the size of avalanches in Fig. \ref{Figslip}(a) exhibits the characteristic power-law \cite{Self-organized_criticality,Earthquakes_as_a_self‐organized_critical_phenomenon,Punctuated_Equilibrium_and_Criticality,Unified_Scaling_law_for_Earthquakes} originally derived for the frequency and size of avalanches for sandpiles by Per Bak {\it et al.} in their theory of Self-Organized Criticality. 
Following the conventions, we identify ``sticks'' as happening at the peaks beyond $F_{1,max}$. The transition from each peak to the next valley is defined as the process of ``slips'', and the change in  $F_1$ will be defined as the avalanche size.
Indeed, an R-square value exceeding 0.9 was obtained when Fig. \ref{Figslip}(b) was fit by a power-law distribution with an exponent of -1.2. 

Now let's concentrate on how $F_{1,{\rm max}}$ changes with $H$. The first equality of Eq.(\ref{EQF_1}) remains valid when $H_c$ is replaced with $H$. Adopting the assumption by Janssen that $\mu_e$ is insensitive to $H$, it predicts $F_1$ to vary as $H^2$ and $H$ for $H\ll A$ and $H\gg A$, which matches the empirical finding in Fig. \ref{Figslip}(c). 

\subsection{Photoelasticity and force chains}

In Secs. IV(A) and IV(B), we emphasize that the existence of a half-buried rod and how much force we pull on it influences the pressure distribution within the granules, with $F_1$ always pointing upward and $F_2$ whose direction may change. This raises two questions: First, how do force chains form and cross-link to support the lifting of granules and the box? Second, what happens when the rod slips, i.e., what determines an unsuccessful lift-off? Photoelastic plates are birefringent commonly used to visualize force chains, meaning that when a polarized light passes through a compressed plate, the polarization direction changes and allows us to observe force chains. By use of such a property, we replaced the granules with two-dimensional photoelastic plates. Unlike the granules, we found shaking less efficient than shearing and compressing to make photoelastic plates more compact.

\begin{figure}
	\includegraphics[width=8.6cm]{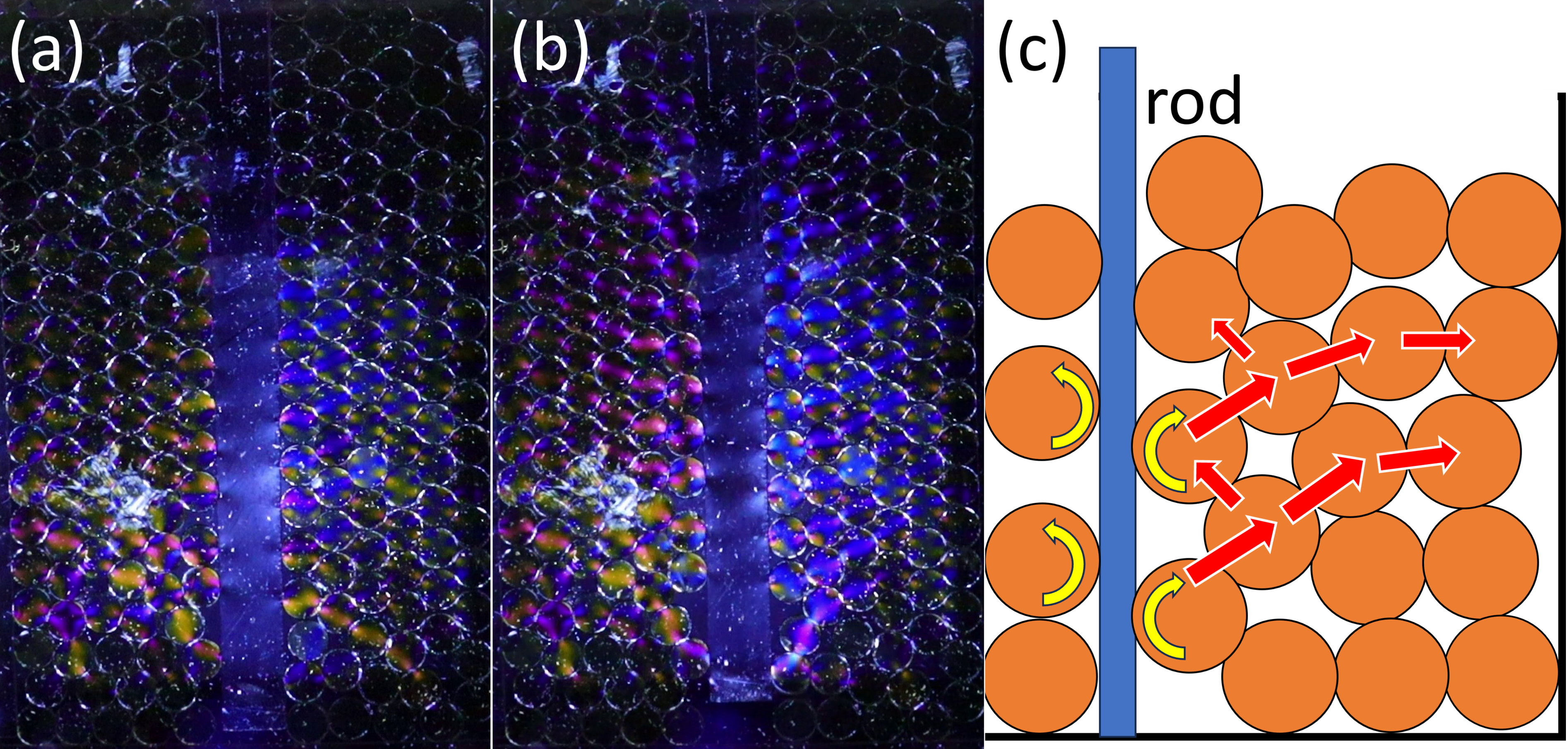}
	
    \caption{(a) Photoelastic fringes reveal the stress field of an acrylic plate half-buried in a 2-D ensemble of photoelastic plates. (b) Fringes change when we successfully lift the whole system by pulling the acrylic plate. (c) A schematic figure demonstrates how the force chains were constructed in the granules under the shearing of a lifted rod.}
    \label{Figphoelas}        
\end{figure}

We expected that force chains evolve from being random without the pull force to fully connecting from the rod to the wall for a successful lift-off. In Fig. \ref{Figphoelas}(a) and Supplementary video 2  Failed to locate the original video, we demonstrated the fun experiment in the Supplementary Video., force chains were scattered and disordered in orientation, which made it hard to connect them and draw a clear network. As soon as the pull force was employed, force chains formed at the interface between photoelastic plates and the rod, which eventually connected to the inner wall of the container. Force chains on the right of the rod are roughly linear with a positive slope in Fig. \ref{Figphoelas}(b), indicating that the directions of $F_1$ and $F_2$ are opposite.  

Experimentally, photoelastic plates would rotate while simultaneously compressing and lifting their neighbors as we pulled the rod, as shown schematically in Fig. \ref{Figphoelas}(c). If the lift-off is unsuccessful, a major avalanche would ensue and destroy all force chains in one fell swoop, causing the value of $F_1$ to fall as in Fig. \ref{Figslip}(a).
Force chains are a manifestation of friction and normal forces, both of which increase during the shearing process. For the rod to successfully lift the whole system, we expect the horizontal pressure on the rod to be much greater than the vertical one to account for the large static friction force. 
When the granules are loosely packed, force chains are prone to collapsing at points where they are not linked to the wall. In the supplementary video, the intensity of optical fringes, which reflects the magnitude of the stress field, is found to weaken near the wall. This is due to the reconstruction of force chains as they relax while the system remains static macroscopically levitated in the air.

\section{Conclusion and discussions}
The classic Janssen effect which describes the pressure distribution inside a granular system was revisited. Our work was motivated by an overheard demonstration \cite{youtube1} in which a half-buried chopstick was used to lift a bottle of rice off the table. 
This feat is counter-intuitive at first glance because, if the friction force provided by the wall of the container is only able to alleviate part of the granular weight according to Janssen, how could the chopstick manage to support the whole weight of granules with a much smaller surface area? 

Combining experiments and models, we clarified the physics overlooked when contesting the viability of such a feat. The main message of our conclusions is that the presence of objects, say, a rod, a tree, or a building, will alter the force chains inside granules - even more so if the objects are under an external force from a human pull or a high wind. Simulating this situation with a simple half-buried rod, we found that the friction force $F_2$ on the wall may switch from originally pointing upward as described by Janssen to downward. This change in direction is accompanied by a dramatic reshuffle of the structure of force chains and reorganizes the pressure distribution in favor of the rod, meaning, the rod receives a much larger normal force than the inner wall to compensate for the former's small area.  

To help understand our experimental findings, theoretical models based on Janssen's approach were attempted to successfully (1) explain the existence of a critical depth of granules for a successful lift-off and predict its value, and (2) elucidate how $F_2$ reverses its direction when $F_1$ increases and why it is crucial to pull off the feat. The same theoretical approach was generalized to describe the case of an unsuccessful lift-off. Specifically, we managed to (3) identify that the occurring frequency for different avalanche sizes obeys a power-law distribution, characteristic of the stick-slip phenomenon. (4) explain how the largest friction force before slipping depends on the initial rod length under the granules.  Unlike the predictions in (1, 3, 4), (2) is hard to verify directly by measurements. Therefore, we resorted to employing the photoelasticity experiment. Indeed, we observed that force chains increase in number and interlink as the pull force increases. This simple experiment provides a direct and heuristic demonstration of why the thin rod could sustain the weight of the whole system, i.e., how the pull force overcame the plight of a small contact area on behalf of the rod by incurring a large normal force on the rod. 

Having mentioned that Furuta {\it et al.} \cite{Packing-dependent_granular_friction_exerted_on_rod} was the first to study such a setup, it is worth emphasizing what our work differs from theirs. Perhaps unaware of the possibility of a successful lift-off, they focused on how a large granular density enhanced the friction force on the rod and ascribed this effect to the increase in pressure due to shear-induced solidification. 
Besides the most interesting properties of this work in (1, 2), our findings in (3, 4)  uncover new aspects of an unsuccessful lift-off. Relevant to their experimental and simulation work, (4) added theoretical models and calculations for the largest friction force. 

This experiment reminded us of the popsicle under the simmering summer heat. The possibility of the wooden stick failing to hold the confection also occurs in our setup, i.e., the whole system suddenly loses its grip on the rod and falls to the table. This sent a strong message that the structure of force chains can experience aging and get weakened over time. Analogous to stress relaxation, the detailed mechanism for the combined system of rod/granules or tree/soil is presumably of practical importance. In other words, this calls for the generalization of the Janseen approach to include dynamics for the pressure distribution.
Preliminary observations in the photoelasticity experiment indeed revealed a reconstruction and decay in force chains, particularly in the horizontal direction, after a successful lift-off. 

Our ongoing efforts include (i) shifting the rod systematically away from the center of the container to force the consideration of a horizontally varying pressure to test one of the core assumptions of Janssen that omits it, 
(ii) reversing the role of the container wall and the rod by detaching the bottom of the container from its wall and gluing the rod to the bottom. This allows the lifting force, now on the container wall, to propagate through the granules through force chains to raise the rod with the bottom of the container and the granules it supports. The first guess is that $H_c$ should be lower for the same experimental parameters since the pulling force is now accompanied by the larger area of the wall. A spoiler: preliminary observations and theoretical calculations similar to those in Sec. III point otherwise. 
So far, the theoretical calculation in Sec. III is still based on the assumptions of Janssen's approach.

\section*{Acknowledgement}
We are grateful to Chung-Hao Chen and Kiwing To for useful discussions, and to the National Science and Technology Council in Taiwan for financial support under Grants No. 112-2112-M007-015 and 113-2112-M-007-008.

\bibliographystyle{plain}

\end{document}